 \documentstyle[11pt,aaspp4]{article}

 \newcommand{\um}{\mbox{$\mu{\rm m}$}}
 \newcommand{\us}{\mbox{$\mu{\rm s}$}}

 \newcommand{\llangle}{\langle}
 \newcommand{\rrangle}{\rangle}
 \renewcommand{\hat}{\widehat} 
 \newcommand{\hhat}[1]{\widehat{#1}}

\begin{document}

 \slugcomment{}

 \title{Fringe visibility estimators for the Palomar Testbed Interferometer}

 \author{M. M. Colavita}

 \affil{Jet Propulsion Laboratory, California Institute of Technology \\
 4800 Oak Grove Dr., Pasadena CA 91109;   Mark.Colavita@jpl.nasa.gov}


 \begin{abstract}

Visibility estimators and their performance are presented for use with the Palomar Testbed Interferometer (PTI).  One operational mode of PTI is single-baseline visibility measurement using pathlength modulation with synchronous readout by a NICMOS-3 infrared array.  Visibility is estimated from the fringe quadratures, either incoherently, or using source phase referencing to provide a longer coherent integration time.  The visibility estimators differ those used with photon-counting detectors in order to account for biases attributable to detector offsets and read noise.  The performance of these estimators is affected not only by photon noise, but also by the detector read noise and errors in estimating the bias corrections, which affect the incoherent and coherent estimators differently.  Corrections for visibility loss in the coherent estimators using the measured tracking jitter are also presented.

 \end{abstract}

 \section{Instrument configuration}

The Palomar Testbed Interferometer (PTI) (\cite{PTInew}; \cite{kent}) uses coherent fringe demodulation and active fringe tracking, similar to that  employed with the Mark~III Interferometer (\cite{M3OI}).  Differences arise attributable to the use of an infrared array detector with its attendant read noise and required bias corrections. 

The beam combiner on PTI accepts the tilt-corrected, delayed beams from the two interferometer apertures.  These are combined at a beamsplitter, and the two combined outputs directed to an infrared dewar.  One output is imaged onto a single pixel of a NICMOS-3 infrared array.  This white-light pixel is band-limited by an astronomical K (2.00--2.40~\um\ FWHM) filter, yielding an effective wavelength of $\sim$2.2~\um.  The other output is dispersed with a prism spectrometer and imaged adjacent to the white-light pixel onto the same line of the infrared array.  Resolution is variable; one typical configuration uses 7 spectrometer pixels with center wavelengths of 1.993--2.385~\um, yielding average channel widths of 65~nm.  The combined light for the spectrometer channels is spatially filtered prior to dispersion with a single-mode infrared fiber.  The white-light channel is not explicitly spatially filtered, although some filtering occurs because of the finite pixel size (40~\um\ pixel and an f/10 relay).

 \section{Array readout}

The infrared array is read out coherently using a 4-bin algorithm with pathlength modulation implemented on the optical delay line. The 100-Hz modulation uses a sawtooth waveform, and the array readout timing varies according to the wavelength of each pixel to achieve a one-wavelength scan for all channels.  Clocking constraints and overhead lead to a typical sample integration time of 6.75~ms (out of a sample spacing of 10~ms) for the white-light pixel, scaling proportionally for other wavelengths. 

For each sample period, the active and adjacent lines of the array are first cleared, the reset pedestal for each data pixel is read, and each pixel is then read out after each quarter-wave of modulation.  Each of these (nondestructive) ``reads'' is actually an average of 16--64 consecutive 2-\us\ subreads, used to reduce the effective read noise, typically to a correlated-double-sample (cds) value of 12~e$^-$ for the white-light pixel and 16~e$^-$ for the spectrometer pixels.  These 5 reads per sample for the white-light and spectrometer pixels are the fundamental interferometer data. 

Denote these 5 reads as $z_i, a_i, b_i, c_i,$ and $d_i$, where $i = 0$ denotes the white-light pixel and $i = 1 \ldots R$ denote the $R$ spectrometer pixels.  The integrated flux in each quarter-wave time bin is calculated as $A_i = a_i -  z_i, B_i = b_i - a_i, C_i = c_i - b_i,$ and $D_i = d_i - c_i.$  From these values, the raw fringe quadratures and total flux are calculated as 
 \begin{eqnarray}
 X_i & = & A_i - C_i \\
 Y_i & = & B_i - D_i \\
 N_i & = & A_i + B_i + C_i + D_i.
 \end{eqnarray}
The total flux $N_i$ (as well as the integrated flux per bin) is related to the actual number of detected photoelectrons $n_i$ by $N_i = kn_i$, where $k$ is a dimensionless gain factor.  For the PTI array electronics, $k$ is typically 0.11.

We can also calculate an energy measure which we denote as
 \begin{equation}
 {\rm NUM}_i  = X_i^2 + Y_i^2.
 \label{eq:NUM}
 \end{equation}
 From these quantities we can estimate the fringe phase, visibility, and  signal-to-noise ratio, but first it is necessary to correct for biases associated with the detection and readout process.

 \section{Biases}

There are several bias terms that need to be measured.  The first set of biases  are the zero points of $A, B, C,$ and $D$, and are those values observed  with the instrument pointing at dark sky.\footnote{With an ideal detector, these biases would be identical, proportional to the dark current and background.  In practice, the biases on $A, B, C,$ and $D$ are slightly different.}(At high light levels, there are also nonlinearities as the detector saturates, but these effects are small for typical observations.) Expressed in terms of the  quadratures and flux, we denote the biases as $B^X, B^Y,$ and $B^N$, so that the corrected values of these quantities are given as (omitting subscripts for  clarity) 
 \begin{eqnarray}
 \hat{X} & = & X - B^X \\
 \hat{Y} & = & Y - B^Y \\
 \hat{N} & = & N - B^N.
 \end{eqnarray}
 We can also correct NUM for these biases as 
 \begin{equation}
 {\rm NUM}^* = {\rm NUM} - B^X(2\hat{X}+B^X) - B^Y(2\hat{Y}+B^Y).
 \end{equation}
 This is equivalent to simply computing ${\rm NUM}^*$ as $\hat{X}^2 + \hat{Y}^2$.

The second set of biases occur in quadratic expressions like ${\rm NUM}$ and arise from the squaring of the photon and read noise.  The two terms are just the variances
 \begin{equation}
 B^{\rm pn} = k\hat{N}
 \label{eq:Bpn}
 \end{equation}
 and
 \begin{equation}
 B^{\rm rn} = 4k^2\sigma_{\rm cds}^2.
 \label{eq:Brn}
 \end{equation}
 Equation~\ref{eq:Bpn} is just the standard photon-counting bias. In Eq.~\ref{eq:Brn}, $\sigma_{\rm cds}$ is the detector read noise (correlated-double-sample), measured in the same units as the integrated flux.  The factor of 4 arises from the 4 bins used to compute NUM.  We are usually read-noise limited on the channels of interest, in which case $B^{\rm rn}$ dominates.   Correcting ${\rm NUM}^*$ for these two variances in addition to $B^X$, $B^Y$, and $B^N$ yields 
 \begin{equation}
 \hhat{\rm NUM} = {\rm NUM}^* - B^{\rm pn} - B^{\rm rn}.
 \end{equation}
 \section{Bias measurements} \label{sec:biasmeasurements}

The biases for each pixel are measured at the beginning of each night of  observation.  While these initial values are adequate for proper operation of  the real-time system, the biases are also measured  repeatedly throughout the night for use in the science data processing. 

 \paragraph{Initial calibrations}

A low-level calibration measurement is made at the beginning of  each night with the instrument pointed at dark sky.  The bias terms $B^X,  B^Y,$ and $B^N$ are computed simply as the measured values of $X, Y,$ and $N$.  The bias term $B^{\rm rn}$ is computed as the mean value of ${\rm NUM}^*$.   This term also incorporates that fraction of the photon-noise bias  attributable to finite dark count and background. 

A high-level calibration measurement is also made at the beginning  of each night using an internal white-light source, which illuminates the white-light and spectrometer pixels.  The increased value of ${\rm  NUM}^*$ with light level is used to estimate the pixel responsivity as 
 \begin{equation}
 k = \left ( {\rm NUM}^* - B^{\rm rn} \right ) / \hat{N},
 \end{equation}
 so that $B^{\rm pn}$ can be computed for other light levels using  Eq.~\ref{eq:Bpn}.  These values of $B^X$, $B^Y$, $B^N$, $B^{\rm rn}$, and $k$ are used by the real-time system.

 \paragraph{On-going calibrations}

Repeated measurements of the bias terms throughout the night accommodate  drifts and improve the quality of the final data processing. Each typically  125-s scan on a science object is bracketed by several other calibration measurements: total-flux {\it foreground} and single-aperture {\it ratio}  calibrations precede the scan; a {\it background} calibration, typically 25-s long,  follows it. 

A foreground measurement observes the target with the instrumental pathlengths intentionally mismatched to yield zero fringe contrast.  In this case, the observed value of ${\rm NUM}^*$ can be used as a direct estimate of the sum $B^{\rm pn} + B^{\rm rn}$. The foreground calibration can also be used to estimate $B^X$ and $B^Y$. 

A ratio calibration measurement observes the target with one aperture  blocked.  Combined with the total flux measured above, the intensity ratio between the interferometer arms can be estimated. 

A background measurement is essentially a low-level calibration measurement  taken close in time to the stellar observation, and as such provides an  estimate of $B^X, B^Y, B^N,$ and $B^{\rm rn}$. 

These five calibration types can be used in different ways in the final data  analysis.  Typically, the biases $B^X$, $B^Y$, $B^N$, and $B^{\rm rn}$ for each scan are  estimated from the associated background measurement, while $B^{\rm pn}$ is calculated from the actual flux during the scan using Eq.~\ref{eq:Bpn}.  Averaging of several nearby background measurements using a median filter generally improves the calibration quality. The current data processing pipeline normally uses the foreground and ratio values only as diagnostics.

 \section{Incoherent estimators}

Given the bias-corrected values $\hat{X}, \hat{Y}, \hat{N},$ and $\hhat{\rm  NUM}$ for the white-light and spectrometer channels, we can estimate fringe visibility. (Strictly, we estimate the square of the amplitude of the complex fringe visibility).  Below we adopt a nomenclature for time intervals: a {\it scan} is a single measurement of an  astronomical target, typically 120--150~s of recorded data, accompanied by  local calibration measurements as described above.  A scan is divided into  {\it blocks}, typically 25~s in length; the fluctuations of estimators among  the blocks of a scan provides an estimate of their internal errors.  Each block comprises a number of {\it frames}, which are typically 0.5~s long, and  synchronized to the half-second tick.  The significance of a frame is that intentional fringe hops to correct unwrapping errors in the real-time system are introduced only at frame  boundaries.  Each frame consists typically of up to 50 {\it samples}, which  are data at the fastest rate in the system, typically 10~ms, which is of order of the atmospheric coherence time.   The actual number of samples  per frame will be less than 50 if fringe acquisition or loss occurs mid-frame;  partial frames with less than typically 10 samples are discarded in the data  processing. 
  Squared visibility $V^2$ is estimated for each channel as (see \cite{vispaper})
 \begin{equation}
 V^2 = \frac{\pi^2}{2} \frac{ \llangle \hhat{\rm NUM} \rrangle}{ \llangle  \hat{N} \rrangle^2},
 \label{eq:V2}
 \end{equation}
 where $ \llangle \rrangle$ represents an average over a  block.\footnote{With step, rather than fringe-scanning modulation, the leading coefficient of Eqs.~\ref{eq:V2}, \ref{eq:compositeV2}, and  \ref{eq:cohV2} would be 4.0.}
 While we're usually not photon-noise limited, the photon-noise-limited SNR is estimated similarly as
 \begin{equation}
 {\rm SNR}^2 = 2\frac{ \llangle \hhat{\rm NUM} \rrangle}{\llangle \hat{N}  \rrangle }.
 \label{eq:SNR2}
 \end{equation}
 The fringe phase is estimated as 
 \begin{equation}
 \phi = \tan^{-1}{\frac{\hat{Y}}{\hat{X}}},
 \end{equation}
 where we make no attempt to be rigorous with respect to phase offset.  These estimates can be made for each channel:  we typically use the suffix  {\it wl} to refer to the white-light channel, viz.\ $V^2_{\rm wl} = V^2_0$.   For the spectrometer channels $1 \ldots R$, we also compute a composite spectrometer visibility $V^2_{\rm spec}$ as 
 \begin{equation}
 V^2_{\rm spec} = \frac{\pi^2}{2} \frac{\sum_i  \llangle \hhat{\rm NUM}_i \rrangle W_i}{\sum_i   \llangle N_i \rrangle^2 W_i}.
 \label{eq:compositeV2}
 \end{equation}
 The range of the summation covers channels $1 \ldots R$, or a subset (for example, $2 \ldots (R-1)$, which excludes the lower-flux channels at the band  edges).  The weights $W_i$ can be uniform, but are typically computed as  $W_i = N_i^2/\sigma_{i,{\rm cds}}^4$, which are proportional to $1/\sigma^2_{V^2_i}$,  as discussed below.  This composite estimator provides an improved signal-to-noise ratio, and is useful for compact sources where visibility changes with  wavelength are smaller than the estimator noise.  However, it still retains  the wide fringe envelope (and thus decreased sensitivity to visibility errors  caused by fringe-tracking errors) corresponding to the narrow spectral  channels of the spectrometer; the use of the weights is useful for accommodating  occasional spectrometer pixels with large read noises.  For consistency, when  the composite visibility is used for science, a composite wavelength computed  with the same weighting is also employed.  At the block level, the SNR of the  $V^2$ estimates is usually sufficiently high that the final $V^2$ estimate for  the scan is calculated as a simple average of the block $V^2$ values, rather  than carrying numerator and denominator separately. 

 \section{Coherent estimators}

We refer to the previous estimators as incoherent, in that NUM, the sum of the square of the fringe quadratures, is computed and summed over the 10-ms  samples; these are generally our default estimators.  However, the complex fringe visibility can be represented by a phasor; if the fringe phase is stable, we can add the phasors vectorially over multiple samples before computing NUM and related quantities.  This ``coadding'' can provide an improved signal-to-noise ratio (\cite{ARAA}),  but at the expense of some atmospheric bias.  To coadd the fringe phasors requires a phase  reference, for which we use the white-light phase $\phi_{\rm wl} = \phi_0$.

We can compute a coherent visibility as follows:  the white-light phase is  scaled by the wavelength ratio between the white-light channel and the channel of interest  to yield $\theta_i = \phi_{\rm wl} \lambda_{\rm wl} / \lambda_i$. The fringe  quadratures are derotated and averaged over $L$ samples as
 \begin{eqnarray}
 (\hat{X}_i)_{\rm coh} & = &  \frac{1}{L}\sum (\hat{X_i}\cos\theta_i - \hat{Y_i}\sin\theta_i ) \\
 (\hat{Y}_i)_{\rm coh} & = &  \frac{1}{L}\sum (\hat{X_i}\sin\theta_i + \hat{Y_i}\cos\theta_i ),
 \end{eqnarray}
 At PTI, the coadd time is typically one 0.5-s frame ($L \simeq 50$ samples), although this is  convenient rather than fundamental. A coherent value of NUM is computed as 
 \begin{equation}
 (\hhat{\rm NUM})_{\rm coh} = (\hat{X})_{\rm coh}^2 + (\hat{Y})_{\rm coh}^2 - (B^{\rm pn} + B^{\rm rn}) / L,
 \label{eq:NUMcoh}
 \end{equation}
 where $L$ reduces the bias correction to account for the reduced noise in the coadded quantities.   From $(\hhat{\rm NUM})_{\rm coh}$ and $(\hat{N})_{\rm coh} =  \langle \hat{N}  \rangle $ for each frame, the coherent $V^2$ is estimated as 
 \begin{equation}
 (V^2)_{\rm coh} = \frac{\pi^2}{2} \frac{ \llangle (\hhat{\rm NUM})_{\rm coh}  \rrangle } { \llangle (\hat{N})_{\rm coh} \rrangle ^2}. 
 \label{eq:cohV2}
 \end{equation}
 A composite $V^2$ for the spectrometer channels can also be computed as for the incoherent case, similar to Eq.~\ref{eq:compositeV2}.

Given that the white-light channel has a high SNR, as required for real-time tracking, the coherent white-light $V^2$ is not an  improved estimator because of coherence losses which occur in the phase-referencing process.  However, it is valuable as an estimator of at least part  of this coherence loss.  We can estimate the coherence loss $\Gamma^{\rm a}$ as
  \begin{equation}
  \Gamma_i^{\rm a} \simeq (V^2_{\rm wl})_{\rm coh} / V^2_{\rm wl},
  \end{equation}
 and we usually divide the coherent spectrometer $V^2$ values through by this value as a partial correction.  To be more exact, one can account for the wavelength difference between the white-light and spectrometer channels by scaling the correction with wavelength as
 \begin{equation}
 \Gamma_i^{\rm a} = \exp \left ( \left (\frac{\lambda_{\rm wl}}{\lambda_i} \right)^2
   \ln \left ( \frac{(V^2_{\rm wl})_{\rm coh}}{V^2_{\rm wl}} \right ) \right ),
 \end{equation}
 which assumes a simple exponential form for the coherence loss. We note that there are additional coherence losses in phase referencing, some of which are discussed in Sec.~\ref{sec:jitter}.

 \section{SNR of the $V^2$ estimators}

The ``detection'' noise on the $V^2$ estimator attributable to photon and read  noise (as opposed to noise attributable to atmospheric turbulence) is readily  calculated.  As is usual, we model only noise on NUM, given by  Eq.~\ref{eq:NUM}, and ignore the smaller noise in $N$ that normalizes  NUM in calculating $V^2$ (\cite{Tango}). The quadratures $X$ and $Y$ are each  comprised of two correlated-double-sample reads, so that the variances of $X$  and $Y$ are given as $\sigma^2_X = \sigma^2_Y = 2\sigma_{\rm cds}^2$. For additive Gaussian noise, $\sigma^2_{X^2} = 2\sigma^4_X = 8\sigma_{\rm cds}^4$, and  similarly for $Y$, yielding $\sigma_{\rm NUM} = 4\sigma_{\rm cds}^2$. Thus,  the standard deviation of the (incoherent) $V^2$ estimate in the read-noise limit is
 \begin{equation}
 \sigma_{V^2} =  \frac{2 \pi^2}{\sqrt{M}}  \left(\frac{\sigma_{\rm cds}}{N}\right)^2, N \ll N_{\rm rn},
 \label{eq:varV2}
 \end{equation}
 where $M$ is the total number of samples, both temporal and spectral,  in the estimate, and is thus applicable to both single-channel and  composite (with equal weights) visibility estimates.\footnote{With step  modulation, the leading coefficient of Eqs.~\ref{eq:varV2}, \ref{eq:varV2coh},  \ref{eq:epsV2}, and \ref{eq:epsV2coh} would be 16.0, with similar changes to Eq.~\ref{eq:Tango}.}
 For arbitrary photon fluxes, read noise can be incorporated into the standard  (4-bin) photon-counting result (\cite{Tango}), yielding
 \begin{equation}
 \sigma^2_{V^2} = \frac{\pi^4}{4MN^4}\left(N^2 + \frac{4}{\pi^2}N^3V^2 +  16\sigma^4_{\rm cds} \right).
 \label{eq:Tango}
 \end{equation}
 Thus the read-noise limit applies when $N \ll N_{\rm rn}$, where
 \begin{equation}
 N_{\rm rn} = \min \left ( 4\sigma^2_{\rm cds}, 3.4V^{-2/3}\sigma^{4/3}_{\rm cds} \right ).
 \end{equation}

A numerical example is illustrative.  For the case of a read noise of 16~e$^-$  per pixel, 125~s of data at 10~ms per sample, and 5 spectrometer channels  in the spectral composite, a standard deviation of 0.02 requires 32 photons  per channel per sample. 

For the coherent estimators, the standard deviation is similar.  Assume as above that $M$ is the total number of 10-ms samples in the  estimate, but that they are first coadded to frames of length $L$ before  calculating NUM.  In this case \begin{equation}
 \sigma_{(V^2)_{\rm coh}} =  \frac{2 \pi^2}{\sqrt{ML}}  \left(\frac{\sigma_{\rm cds}}{N}\right)^2, N \ll N_{\rm rn},
 \label{eq:varV2coh}
 \end{equation}
 Thus, the required photon flux for a given accuracy scales as $L^{- 1/4}$.  With $L$ = 50 and the parameters above, an accuracy of 0.02 now requires 12 photons per channel per sample, although, as discussed above, the coherent estimate is more susceptible to systematic biases.
 
 \subsection{SNR for bias estimation}

Strictly speaking, the above analysis is somewhat simplistic, as it assumes  that bias correction adds no additional noise.  For low light levels, the largest errors in bias correction are attributable to estimation of the biases $B^{\rm pn}$ and  $B^{\rm rn}$; errors  in their estimation are additive with the noise on NUM as calculated above.  However, as $B^{\rm pn}$ and  $B^{\rm rn}$ are computed from NUM measured under known  conditions (Sec.~\ref{sec:biasmeasurements}), the $V^2$ errors due to  imperfect bias estimates can be computed using the expressions above.  Thus,  for incoherent quantities, the (incoherent) $V^2$ error due to imperfect bias  subtraction is given by 
 \begin{equation}
 \epsilon_{V^2} \simeq \frac{2 \pi^2}{\sqrt{M_b}}  \left(\frac{\sigma_{\rm cds}}{N}\right)^2,  N \ll N_{\rm rn},
 \label{eq:epsV2}
 \end{equation}
 where $M_b$ is the number of samples used in estimating the bias. This  expression is strictly accurate only for the read-noise bias $B^{\rm rn}$, or  when both $B^{\rm pn}$ and  $B^{\rm rn}$ are computed from a foreground calibration. However at low photon  fluxes, where bias errors are most significant, the read-noise term is  dominant and the above expression is a good approximation. 

For the coherent $V^2$, the situation is somewhat better, as the errors in $B^{\rm pn}$ and  $B^{\rm rn}$ are reduced by the number of samples in the coherent average, per Eq.~\ref{eq:NUMcoh}.  For the biases computed incoherently, and applied via Eq.~\ref{eq:NUMcoh}, the applicable expression is
 \begin{equation}
 \epsilon_{(V^2)_{\rm coh}} \simeq \frac{2 \pi^2}{\sqrt{M_b} L}  \left(\frac{\sigma_{\rm cds}}{N}\right)^2, N \ll N_{\rm rn},
 \label{eq:epsV2coh}
 \end{equation}
 subject to the same caveats at Eq.~\ref{eq:epsV2}. By way of comparison, if the biases were computed ``coherently'', i.e., from measured values of $(NUM)_{\rm coh}$, then Eq.~\ref{eq:epsV2coh} would have the same dependence on $L$ as Eq.~\ref{eq:varV2coh}.

Thus, the total ``detection'' noise on $V^2$ is the quadrature sum of  $\sigma$ and $\epsilon$, and the contribution due to bias estimation can be important.  This contribution is generally not important on bright sources where the noise on $V^2$  is dominated by atmospheric effects.  On fainter targets, the relative  bias noise can be decreased by incorporating additional calibration data (for  example, using background calibrations from a larger time window about the  science scan, rather than just its explicitly-associated background), although  eventual nonstationarity of the underlying statistics presents a practical  limit. 

While calibration errors are usually dominated by the $B^{\rm pn}$ and  $B^{\rm rn}$ terms, the errors attributable to the other bias terms are easily computed: for both incoherent and coherent estimators, the errors $\epsilon^X_{V^2}$, $\epsilon^Y_{V^2}$, and $\epsilon^N_{V^2}$  associated with $B^X$, $B^Y$, and $B^N$ are given by
 \begin{eqnarray}
 \epsilon^X_{V^2} = \epsilon^Y_{V^2} & = & \frac{\sqrt{2} \pi  V}{\sqrt{M_b}}\left ( \frac{\sigma_{\rm cds}}{N} \right ), N \ll N_{\rm rn} \\
 \epsilon^N_{V^2} & = & \frac{2 V^2}{\sqrt{M_b}}\left ( \frac{\sigma_{\rm cds}}{N} \right ), N \ll N_{\rm rn}.
 \end{eqnarray}

 \section{Data quality measures}

Inter-block fluctuations of estimated quantities are useful to estimate internal errors.  However additional data quality measures are available.

 \subsection{Lock time}

PTI uses a multi-stage algorithm for fringe acquisition and track (\cite{PTInew}). Essentially, the average SNR must exceed a given threshold for the system to enter the ``locked'' state; loss of lock and reacquisition occurs if the SNR falls below a second threshold.  Fringe data is only recorded when locked; to account for the time delay caused by the memory of the  averaging filter in detecting loss of lock, data at the end of a lock is automatically expunged. Thus, with multiple locks, the elapsed time to collect a fixed amount of data in order to complete a scan is increased.

Two heuristic data-quality measures are the fraction of lock time to elapsed  time, and the number of separate locks that make up the total data  on a scan.  For bright stars and good seeing, each scan is comprised of just  several long locks.  For very faint stars, or with poor seeing, each scan can be  comprised of many short locks, reflecting the inability of the system to  consistently track the fringe.  While visibility can be estimated in all  cases, the data quality in the latter case will be inferior.  Typically, this  poorer data quality is evident in the inter-block fluctuations, in which  case the lock-time  metric is only advisory. 

 \subsection{Jitter}
 \label{sec:jitter}

We can estimate a first-difference phase jitter $\sigma_{\Delta \phi}$ as
 \begin{equation}
 \sigma_{\Delta \phi}^2 = \llangle \left( \phi_{\rm wl}(n) - \phi_{\rm wl}(n-1) \right)^2 \rrangle,
 \label{eq:jitter}
 \end{equation}
 where $\phi_{\rm wl}$ is computed from the 10-ms samples.  While this quantity is not unbiased with respect to detection noise, successful fringe tracking typically requires an SNR $>$ 5, so that the detection bias on $\sigma_{\Delta \phi}^2$ should be $<$ 0.08.

With an ideal instrument, $\sigma_{\Delta \phi}$ is related to the atmospheric  coherence time.  Coherence time can be defined in various ways (\cite{Buscher}).  Let $t_{0,i}$ denote the structure-function definition of coherence time, viz.\ that sample spacing for which the phase difference between samples is one radian rms.  The structure function depends on the actual sample spacing $t$ as $D_i(t) = (t/t_{0,i})^{5/3}$.  For $i=1$, representing the contribution from a single aperture (the usual adaptive-optics definition), $t_{0,1}$ = $0.314r_0/W$ for coherence diameter $r_0$ and constant wind speed $W$; for $i=2$, representing contributions from two apertures (applicable to interferometry), $t_{0,2} = 0.207r_0/W$.

Let $T_{0,i}$ to denote the variance definition of coherence time, viz.\ that sample integration time for which the phase fluctuations about the interval mean are one radian rms.  It is given by $T_{0,1} = 1.235r_0/W$ and $T_{0,2} = 0.815r_0/W$.  The structure function depends on the actual sample integration time $T$ as $\sigma_i^2=(T/T_{0,i})^{5/3}$.

Thus, for an ideal  instrument, the coherence time $T_{0,2}$ can be estimated as 
 \begin{equation}
 T_{0,2} = 3.94 t / \sigma_{\Delta \phi}^{6/5},
 \end{equation}
 where $t$ is the sample spacing.  Fringe motion during the sample integration time $T$ blurs the fringe, reducing the visibility.  For rapid  (with respect to underlying phase motion) fringe scanning, the coherence reduction is related to the high-pass fluctuations about the interval mean, $(\sigma_{\phi})_{\rm hp}$, as $\Gamma^b = \exp(-(\sigma_{\phi})^2_{\rm  hp})$, or given the coherence definitions above, $\Gamma^b = \exp(- (T/T_{0,2})^{5/3})$. 
 We can write this in terms of the phase jitter as
 \begin{equation}
 \Gamma^b = \exp \left (-C_{\Gamma} \sigma_{\Delta \phi}^2 \right ),
 \end{equation}
 with the coefficient $C_{\Gamma}$ given by
 \begin{equation}
 C_{\Gamma} = \left(\frac{T}{3.94t}\right)^{5/3}.
 \end{equation}
 For $T$ = 6.75~ms (for the white-light pixel) and $t$ = 10~ms,  $C_{\Gamma}$ = 0.053. 

A more careful calculation of $C_{\Gamma}$ can be done for this case (Appendix~\ref{sec:app}), accounting for the finite integration time  required to measure $\phi_{\rm wl}$, which yields $C_{\Gamma}$ = 0.057.  A similar calculation can be done under the assumption that all phase noise is caused by narrow-band vibrations with frequency $\ll 1/t$; in this case, $C_{\Gamma}$ = 0.038.

When we apply this correction to PTI data, we usually err on the side of undercorrection by adopting a modest leading coefficient of 0.04. In general, an empirical  visibility-reduction coefficient can be adopted from fits to the measured data  applicable to the actual atmospheric realization and instrumental configuration.  However, for data calibrated with spatially- and temporally-local calibrators  (and especially if the calibrators are of similar brightness to the target),  the reduction in visibility due to the above temporal effects will be mostly  common mode and divide out of the normalized visibility.  In this case, the  value of the jitter is useful as a measure of the seeing, and indirectly of  the data quality. Finally, we note that the coherent $V^2$ estimates on PTI often exhibit  coherence losses larger than predicted from the models above.  These may be  attributable to different apodizations of the starlight pupil between the spectrometer and white-light sides of the beamsplitter.  In particular, the single-mode fiber preceding the spectrometer imposes a Gaussian  apodization on the pupil, while the white-light channel---with no explicit  spatial filter---imposes a more uniform pupil weighting.  These different apodizations will result in slightly different instantaneous phases between the two beamsplitter outputs, and thus a coherence loss when phase referencing the spectrometer channels to the white-light phase.

\subsection{Ratio Correction} 

PTI uses a single-mode fiber after beam combination to spatially filter the  spectrometer channels.  Spatial filtering increases the raw visibility and  reduces the concomitant noise attributable to spatial effects; temporal  effects must still be calibrated.  As spatial filtering by the fiber essentially rejects light which would not interfere coherently, there is induced scintillation, which has a second-order effect on visibility.  With simultaneous intensity measurements of each arm in a fully single-mode  combiner (\cite{Foresto}), an essentially perfect correction for this effect is  possible, but it can be shown (\cite{Shaklan}) that measurement of only the  average intensity ratio between the two arms is adequate. If we denote this  ratio as $R_{12}$, then the correction for the induced scintillation is
 \begin{equation}
 S_{12} = \frac{(1+R_{12})^2}{4R_{12}}.
 \end{equation}
 As discussed in Sec.~\ref{sec:biasmeasurements}, the combination of the foreground and ratio measurements allows estimation of $S_{12}$ for each scan.

Currently, strict application of the ratio correction at PTI has been unsatisfactory, and we generally do not apply it. We attribute this to two effects.  One is that given noisy values  of $R_{12}$, $S_{12}$ is a biased estimator, and will tend to over-correct the  visibility.  The second is that the measurements of the ratio are not truly  simultaneous with the scan.  Thus, seeing nonstationarity will affect the  estimate.  Also, there is a selection effect as fringe data is only recorded  when locked, while the flux calibrations are contiguous. 

Even without the ratio correction, the spatially-filtered data yields significantly-improved  visibility estimates.  However, the ratio correction has been useful as an  additional indicator of data quality.  For example, at high zenith angles,  asymmetric (due to misalignment) vignetting in the system will increase $S_{12}$.  But as with jitter, vignetting is tied to sky position, and spatially-local  calibration will ameliorate most of the systematic visibility effects. 

\section{Conclusion}

The use of array detectors at PTI requires attention to bias correction in fringe-parameter estimators, especially energy measures like $V^2$ which use squared quantities.  Observations with PTI incorporate nightly and on-going bias calibrations, which can be used to compute optimal bias corrections.  In addition to statistical noise in the estimators themselves, noise in the bias terms plays a role in the overall data quality.  Inter-block fluctuations of estimated quantities are useful to estimate internal errors.  Auxiliary data quality metrics include the tracking jitter and the ratio-correction estimate, which can be used for open-loop corrections or as independent data quality measures.

\acknowledgements

Thanks to Fabien Malbet and Gerard van Belle for useful comments.  The work reported here was conducted at the Jet Propulsion Laboratory, California Institute of Technology, under contract with the National Aeronautics and Space Administration.

 \appendix

\section{Temporal coherence calibration using the phase jitter} \label{sec:app} We start with the assumption that the coherence reduction on $V^2$ can be written as
 \begin{equation}
 \Gamma = \exp \left ( -(\sigma_{\phi})^2_{\rm hp} \right ),
 \label{eq:ap1}
 \end{equation}
 where $(\sigma_{\phi})_{\rm hp}$ is the high-pass phase jitter. This is strictly true for the case where the fringe scanning is much faster than any frequencies of interest, although the results are similar with a slower scan.
 The high-pass jitter in Eq.~\ref{eq:ap1} is given by the frequency-domain integral 
 \begin{equation}
 \sigma^2_{\rm hp} = \int {\rm d}f W(f) (1-\mbox{sinc}^2(\pi f T)),
 \end{equation}
 where $W(f)$ is the phase power spectrum, $1-\mbox{sinc}^2()$ is a high-pass  filter, and $T$ is the sample integration time.  A similar spectral representation exists for the phase jitter (Eq.~\ref{eq:jitter}):
 \begin{equation}
 \sigma^2_{\Delta \phi} = \int {\rm d}f W(f) \mbox{sinc}^2(\pi f T)  4\sin^2(\pi f t), 
 \end{equation}
 where $\mbox{sinc}^2()$ accounts for averaging over the sample  integration time, while $\sin^2()$ is a high-pass filter corresponding to a sample  spacing of $t$. 

For $f<1/T$, the filter function in the integral for $\sigma^2_{\rm hp}$ is $H_{\rm hp}(f) \simeq \frac{1}{3}\pi^2T^2f^2$, while for $f<1/t$,  the filter function in the integral for $\sigma^2_{\Delta \phi}$ is $H_{\Delta \phi}(f)  \simeq 4\pi^2t^2f^2$.  The ratio of the filter functions is 
 \begin{equation}
 C_\Gamma = \frac{H_{\rm hp}(f)}{H_{\Delta \phi}(f)} =  \frac{1}{12}\left(\frac{T}{t}\right)^2. 
 \end{equation}
 With $T$  = 6.75~ms (for the white-light pixel) and $t$ = 10~ms,  we calculate $C_{\Gamma}$ = 0.038. Thus, for narrowband low-frequency noise, we can write  the visibility reduction directly in terms of the first-difference variance as 
 \begin{equation}
 \Gamma^{\rm b} = \exp \left (-C_{\Gamma}\sigma^2_{\Delta \phi} \right ).
 \end{equation}

This same formulation applies for other noise models. For $W(f)$ given by an atmospheric power spectrum, $W(f) \propto f^{-8/3}$ (assuming a low fringe-tracker bandwidth), it is necessary  to compute the integrals numerically. For power laws of the form $f^{- \alpha}$, some representative values of $C_{\Gamma}$ for $T/t$ = 0.675 are 0.057, 0.070, 0.088, and 0.145 for $\alpha$ = 8/3, 2.5, 7/3, and 2.0, respectively.

 \end{document}